Regulate the direct-indirect electronic band gap transition by electron-phonon interaction in BaSnO$_3$

Binru Zhao, Qing Huang, Jiangtao Wu, Jinlong Jiao, Mingfang Shu, Gaoting Lin, Qiyang Sun, Ranran Zhang, Masato Hagihala, Shuki Torri, Guohua Wang, Qingyong Ren, Chen Li, Zhe Qu, Haidong Zhou*, Jie Ma*

ABSTRACT: The neutron powder diffraction, specific heat, thermal conductivity, and Raman scattering measurements were presented to study the interplays of lattice, phonons and electrons of the Sr-doping Ba$_{1-x}$Sr$_x$SnO$_3$ ($x \leq 0.1$). Although Ba$_{1-x}$Sr$_x$SnO$_3$ kept the cubic lattice, the Raman spectra suggested a dynamic distortion at low temperature. The density functional theory was applied to analyze the electronic structures and phonon dispersions of Ba$_{1-x}$Sr$_x$SnO$_3$($x = 0, 0.0125$), and the behaviors of electron bands around Fermi levels were discussed. According to the experimental and theoretical results, the Sr-doping played a significant role in tuning the indirect band gap of BaSnO$_3$ and influenced the electron-phonon interaction.

INTRODUCTION

Due to the novel physical properties, such as superconductivity, charge density wave, ferroelectricity, thermoelectricity and optical absorption,[1-5] the bulk compounds with significant electron-phonon interaction has sparked intense research upsurge in condensed matter physics[6,7]. The charged ions from the equilibrium positions not only unbalance the local charge with the modified electrostatic potential, but also adjust the phonon excitation in the crystal.[8] On the other hand, the increase of charge carriers could drastically improve the electrostatic screening and lower the electron-phonon interaction. Although it has been well recognized that the electron-phonon interaction is strong in the nonmetals, the electron contribution is mainly focused and the related phonon transport has been little investigated.

The perovskite oxides, ABO$_3$, could exhibit a semiconductor behavior with a wide band gap, and induce a strong impact on the lattice of the BO$_6$ octahedral. For instance, Barium titanate, BaTiO$_3$ (BTO), has attracted a lot of attention as a ferroelectric oxide from the off-center Ti$^{4+}$ ions and the ferroelectric distortion has been recognized to be driven by the orbital hybridization between $3d$ states of $d^0$ ion Ti$^{4+}$ and O $2p$ states.[9] However, the localized Ti $3d$ states lead to the heavy electron effective mass, consequently limiting the carrier mobility. If the B-site Ti$^{4+}$ ion was replaced by a large ion, Sn$^{4+}$ ion, the first-principle calculations suggested the barium stannate, BaSnO$_3$ (BSO), to be dominantly contributed by $2s$ states of $s^0$ ion Sn$^{4+}$ with the small effective mass and result in a low electron density-of-states with an indirect gap of 3.1 eV. Meanwhile, BSO is cubic at room temperature, Fig. 1(a).

BSO exhibits a wide doping tunability from the high symmetric lattice structure, while the A-site substitution could adjust the environment of BO$_6$ octahedral and induce lattice distortion to effectively tune the optical property. In order to explore the unique property of BSO, the Sr ion is introduced on the Ba-site. Sr is in the same II$_A$ group as

Ba, while the atomic radius is a trifle smaller than that of Ba atom. $SrSnO_3$ (SSO) is similar as BSO, such as the low electron effective mass[10-12], and the indirect band gap of 4.0 eV[13] suggests enhanced bond energy and the localized electrons. Hence, Sr replacement to Ba cations induce the strong ionic size effect on the band structure and electron-phonon coupling.

In this work, we presented a systematic illustration of the structural and thermal properties of $Ba_{1-x}Sr_xSnO_3$ ceramics (BSSOs, $x$ = 0, 0.03, 0.06, 0.1), viz. the lattice parameters, heat capacity, thermal conductivity and Raman shift, as the function of Sr substitution and temperature. Combining the simulations of the phonon dispersion and the electronic band structures of $Ba_{1-x}Sr_xSnO_3$ ($x$ = 0, 0.125), not only the lattice vibrations and electronic structures had been analyzed in details, but the strong electron-phonon interaction had been discussed.

## RESULTS AND DISCUSSION

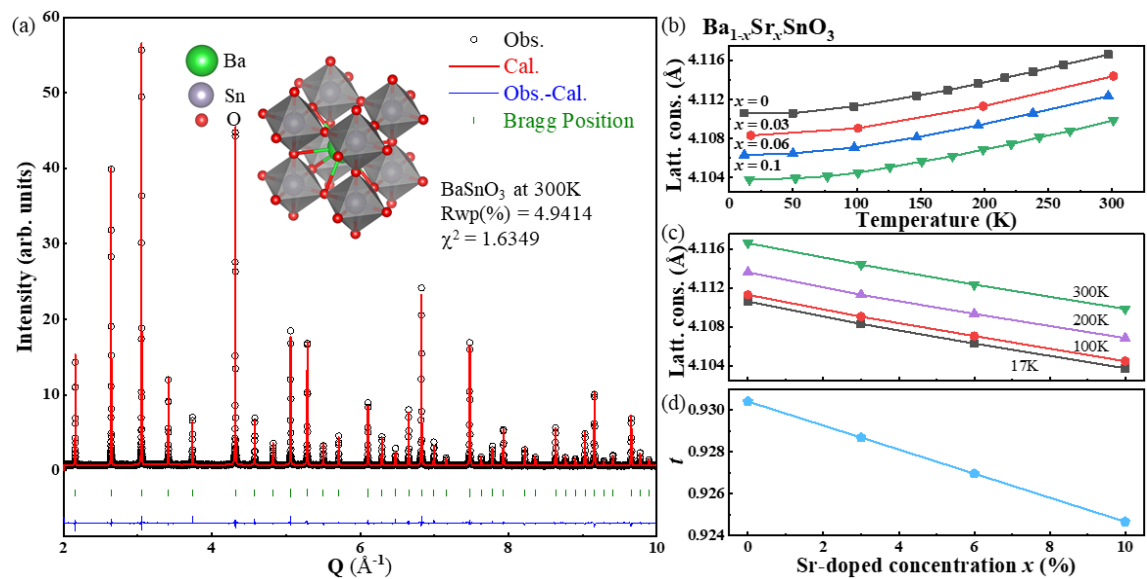

Figure 1. (a) Neutron powder diffraction pattern and simulation of $BaSnO_3$ at 300 K. (b) and (c) Temperature- and Sr-doping-dependence of the lattice parameters of BSSOs, respectively. (d) The Goldschmidt tolerance factor $t$ as a function of Sr-doped concentration at 300 K.

**Neutron Powder Diffraction.** Figure 1 (a) presented the comparison of Neutron powder diffraction (NPD) data and Rietveld refinement of BSO at 300 K with a space group $Pm\bar{3}m$. The lattice parameters as a function of temperature and Sr-doped concentration were shown in Figure 1 (b) and (c), respectively. As the temperature increased, the lattice parameters of BSSO monotonically increased. Meanwhile, the lattice constants decreased with the increase of Sr-content due to the smaller radius of $Sr^{2+}$ ion than that of $Ba^{2+}$ ion. The lowest lattice parameter was 4.104 Å in $Ba_{0.9}Sr_{0.1}SnO_3$ at 17 K.

The Goldschmidt tolerance factor, $t$, was used to predict the stability of the perovskite structure, $ABO_3$,

$$t = \frac{r_A + r_O}{\sqrt{2}(r_B + r_O)} \quad (1)$$

where $r_A$, $r_B$, $r_O$ were the ionic radius.

The ionic radius of $Sn^{4+}$ and $O^{2-}$ ions were 69 and 140 pm, respectively, and the ionic radius on A-site referred to the combination of $Ba^{2+}$ (135 pm) and $Sr^{2+}$ (118 pm) ions. The values of $t$ were obtained by weighted average in doping compounds.[14] Compared to 1 for the ideal cubic structure, $t$(BSO) was 0.93 and the polar distortion or oxygen octahedra tilting was observed. As the smaller Sr-ion substituted the Ba-ion, the lattice distortion increased with a lower $t$ and both bond lengths of (Ba, Sr)-O and Sn-O decrease, which increased the interaction between A-site (Ba, Sr) and Sn cations.

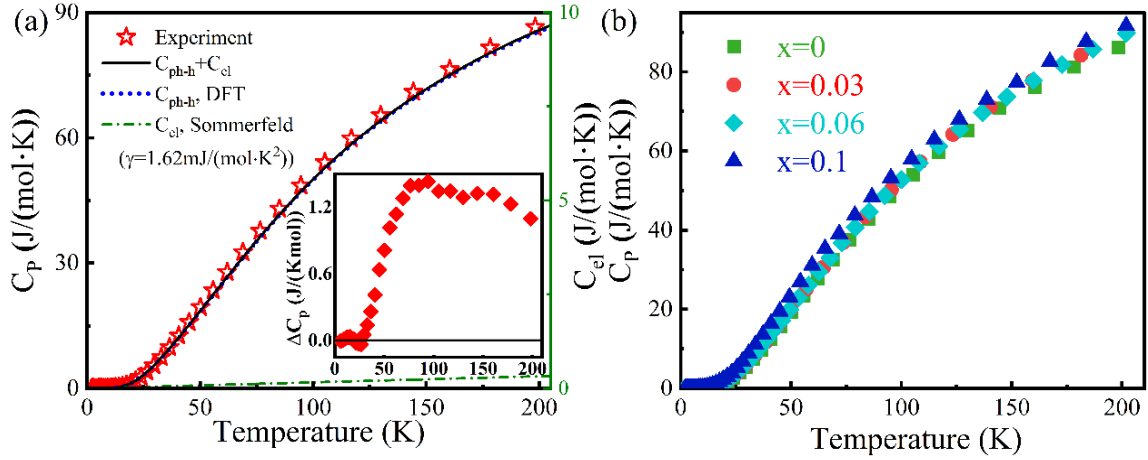

Figure 2. (a) The heat capacity $C_p$ of $BaSnO_3$. The $C_{ph\text{-}DFT}$ of harmonic phonons is calculated by *phonopy* and $C_{el}$ is the electron heat capacity in Sommerfeld approximation. (b) $C_{PS}$ of BSSOs from 2 K to 200 K.

**Specific Heat**. The specific heats, $C_{PS}$, of the BSSOs were measured from 2 to 200 K and agreed with the calculations well, Fig. 2. The electronic specific heat, $C_{el}$, was estimated by the Sommerfeld Theory, *i. e.* $C_{el} = \gamma_{el}T$, where $\gamma_{el}$ was the Sommerfeld parameter as 1.62 mJ/(mol·$K^2$) for BSO.

To investigate the lattice specific heat, $C_{ph}$, we calculated the specific heat of harmonic phonon, $C_{ph\text{-}h}$, by the *phonopy* code on density function theory (DFT) and estimated the anharmonic effect, $C_{ph\text{-}an}$, qualitatively. $C_{ph\text{-}h}$ did not depend on volume and temperature, $C_{ph\text{-}h,V} = C_{ph\text{-}h,P}$, while $C_{ph\text{-}an}$ depended on both factors of volume and temperature. Therefore, the difference of $\Delta C_P = C_P - C_{ph\text{-}h} - C_{el}$ was mainly described as $C_{ph\text{-}an}$, the inset of Fig. 2(a), which was started to be observed around 40 K. Above 100 K, $\Delta C_P$ gradually decreased for the Dulong-petit theory limitation, $3nR$, where $n = 5$ was the atomic number per formula and the molar gas constant $R = 8.31$ J/(mol·K)).

Figure 2(b) was the $C_{Ps}$ of BSSOs from 2 K to 200 K, and the $Sr^{2+}$-content induced to a higher $C_P$ by enhancing the lattice distortion and introducing the impurity-anharmonicity term, $C_{ph\text{-}Im,A}$. Moreover, the $C_P$ difference between Sr-doped BSO and the parent compound was less than 3% at high temperature due to the Dulong-Petit limitation. The lattice dynamics would be discussed in details in the section of the

Raman spectra to explain this anomaly.

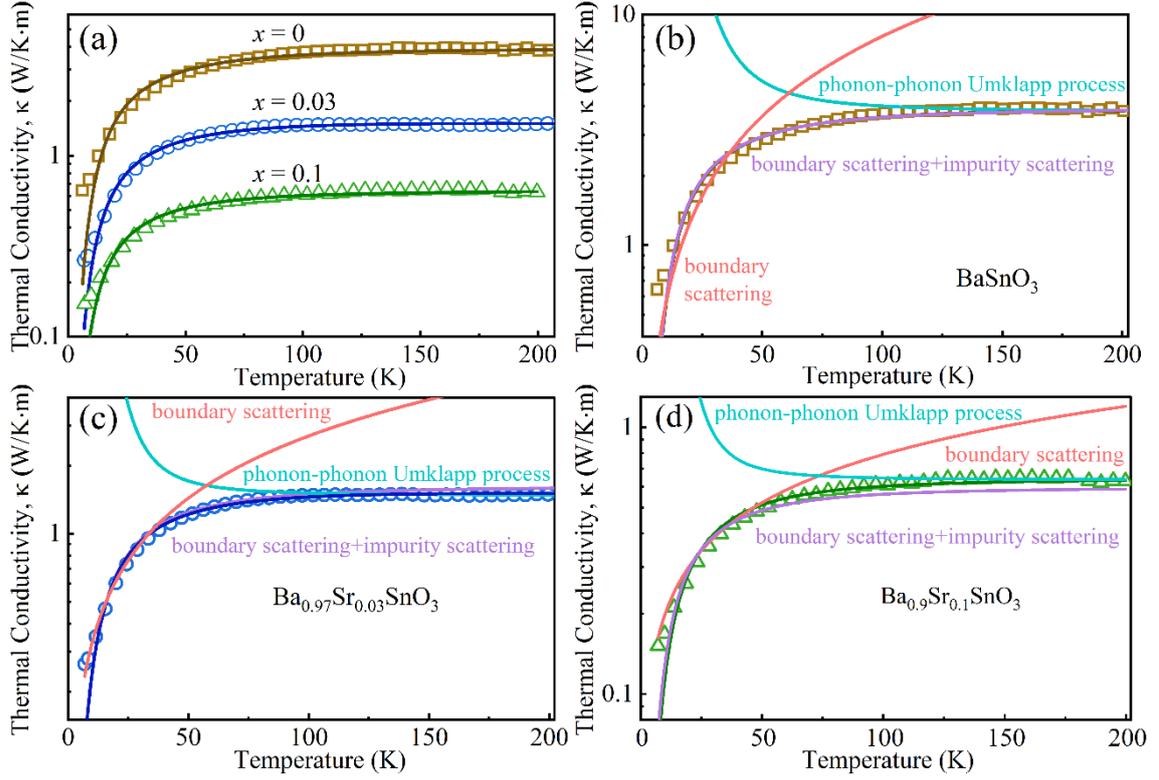

Figure 3. (a) The thermal conductivities κs of Ba$_{1-x}$Sr$_x$SnO$_3$($x$ = 0, 0.03, 0.1) in the temperature range 20 K-300 K. (b)-(d) The thermal conductivities contributed by boundary scattering $\kappa_b \sim T^3$, impurity scattering described by the part of Debye-Callaway model and Umklapp process $\kappa_U \sim A\exp(-\Theta_D/3x)T^{-2}$.

**Thermal Conductivity.** Figure 3 presented the thermal conductivity, κ, of BSSOs. Due to the wide band gap of BSO, [15] κ was mostly contributed by the lattice and was gradually flattened above around 100 K. In order to obtain different contributions of the lattice thermal conductivity, and the nonmagnetic simplified Debye-Callaway model[16] based on Boltzmann distribution was applied,

$$\kappa_{lattice} = \frac{k_B}{2\pi^2 v_{ph}} \left(\frac{k_B T}{\hbar}\right)^3 \int_0^{\theta_D/T} \frac{\tau_{ph} x^4 e^x}{(e^x - 1)^2} dx \qquad (2)$$

$$\frac{1}{\tau_{ph}} = \frac{1}{\tau_U} + \frac{1}{\tau_b} + \frac{1}{\tau_i} \qquad (3)$$

$$\frac{1}{v_{ph}} = \frac{1}{3}\left(\frac{2}{v_t} + \frac{1}{v_l}\right) \qquad (4)$$

where $x = \hbar\omega/k_B T$, and $\theta_D$ was the Debye temperature. $\tau_{ph}$ was the phonon relaxation time which affected by the Umklapp process ($\tau_U$), boundary scattering ($\tau_b$) and impurity scattering ($\tau_i$) including both mass-difference scattering and dislocation scattering. $v_t$, $v_l$, and $v_{ph}$ were the transverse, longitudinal and average sound velocities, respectively. The mean lifetime of boundary scattering was expressed as $\tau_b^{-1} = v_{ph}/L$ with the grain size ($L$) for the polycrystal. The phonon-phonon Umklapp process relaxation time was $\tau_U^{-1} = BT^3\omega^2 e^{-\theta D/3T}$ for the insulators, and both $B$ and $b$ were fitted by the experiment data. The phonon relaxation time of impurity scattering was $\tau_i^{-1} = A\omega^4$. For the Sr-doped BSO, the impurity term was modified by Klemens [17]:

$$A = \frac{V}{4\pi v_s^3} \cdot \Gamma \qquad (5)$$

$$\Gamma = \Gamma_{MF} + \Gamma_{SF} \qquad (6)$$

where $V$ was the volume per atom, $\Gamma$ included the disorder induced by mass difference fluctuation ($\Gamma_{MF}$) and strain field fluctuation ($\Gamma_{SF}$) by dislocation.

The lattice distortion of the different atomic sizes at $A$-site and interatomic force field caused by the host atom would further scatter the phonon.[18]

$$\Gamma_{MF} = \frac{1}{3}x(1-x)\left(\frac{M_{Ba} - M_{Sr}}{M_{aver}}\right)^2 \qquad (7)$$

$$\Gamma_{SF} = \frac{1}{3}x(1-x)\varepsilon\left[\frac{(1-x)M_{Ba} + xM_{Sr}}{M_{aver}}\right]^2 \left[\frac{r_{Ba} - r_{Sr}}{xr_{Sr} + (1-x)r_{Ba}}\right]^2 \qquad (8)$$

where $M_{Ba}$, $M_{Sr}$, $M_{aver}$ represented the masses of Ba and Sr ions and the average ionic mass in one primitive cell, respectively. $x$ meant the substituted fraction, $\varepsilon$ was the adjustable parameter in the range of 10 to 100 for the Sr sublattice[18], $r_{Ba}$ and $r_{Sr}$ were the atomic radius of Ba and Sr, respectively.

As shown in Fig. 3(b)-(d), the crystal grain boundary mainly scattered the phonons at low temperatures, while the contribution of impurity should be included with the increasing temperature. Finally, the U-process flattened the curve at high temperatures by reinforcing the anharmonic phonon-phonon interaction.

**Table 1. $\Gamma_{MF}$, $\Gamma_{SF}$ and ε were combined the impurity scattering coefficient and resulted in the parameter A in Eq.5. B was the Umklapp scattering coefficient and b was the Umklapp scattering temperature coefficient. v/L represented the boundary scattering.**

| Configuration | $\Gamma_{MF}$ | $\Gamma_{SF}$ | ε | A($s^3$) | B($s^2$/K) | v/L(m/$s^2$) |
|---|---|---|---|---|---|---|
| BaSnO$_3$ | - | - | - | 6.17×10$^{-42}$ | 1.6×10$^{-33}$ | 8.73×10$^9$ |
| Ba$_{0.97}$Sr$_{0.03}$SnO$_3$ | 0.01 | 0.10 | 84.80 | 1.91×10$^{-41}$ | 9.9×10$^{-32}$ | 2.45×10$^{10}$ |
| Ba$_{0.9}$Sr$_{0.1}$SnO$_3$ | 0.03 | 0.24 | 69.57 | 6.25×10$^{-41}$ | 1.0×10$^{-33}$ | 5.70×10$^{10}$ |

As the Sr concentration increased, the thermal conductivity was reduced. The disorder coefficients $\Gamma_{MF}$ and $\Gamma_{SF}$ were listed in Table 1: the replacement of the lighter/smaller Sr ions not only leaded to the interaction between Ba and Sr atoms on Eq. (7) by the mass difference, but also enhanced the lattice distortion by the strain field including the SnO$_6$ octahedra. Additionally, the contribution from mass difference and strain field were variable in doped compounds. For Ba$_{0.97}$Sr$_{0.03}$SnO$_3$ configuration, $\Gamma_{MF}$ and $\Gamma_{SF}$ had same magnitude, suggesting the comparable contributions from the mass difference and strain field. However, $\Gamma_{MF}$ was significantly larger than $\Gamma_{SF}$ in the doped compounds. Which we mentioned above demonstrated that the lattice distortion from the strain field played an important role to scatter phonons due to the Sr substitution. Above around 100 K, the combination of impurity scattering, phonon-phonon scattering, phonon-boundary scattering and potential electron-phonon coupling simultaneously affected the lattice thermal conductivity and introduced the flat κ.

**Table 2. Frequencies (cm$^{-1}$) of Raman peaks in experiment, compared with the calculated phonon modes of BaSnO$_3$ at the high symmetry points R, M, Γ and in Pbnm structure. The star symbol * represented the second-order Raman mode.**

| Raman data | Phonon mode | R$\bar{3}$c R (0.5,0.5,0.5) | P4/mbm M$_1$ (0.5,0.5,0) | Pbnm | Γ (0,0,0) |
|---|---|---|---|---|---|
| 266 | TO$_2$ |  |  | 260 | 248 |
| 278 | TO$_2$ | 275 |  |  |  |
| 286 | TO$_2$ |  | 288 |  |  |
| 299 | LO$_2$ |  |  | 292 |  |
| 357 | LO$_2$ | 344 |  |  | 384 |
| 479 | 2TO$_2$* |  |  |  |  |
| 527 | TO$_3$ |  | 522 |  |  |
| 553 | TO$_3$ |  |  | 553 |  |
| 595 | TO$_3$ |  |  | 579 |  |
| 610 | TO$_3$ |  |  | 596 |  |
| 629 | TO$_3$ |  |  |  | 645 |
| 657 | TO$_3$ |  | 656 |  |  |

**Raman Spectra.** To further investigate the intrinsic phonon behavior, the temperature dependent Raman spectra of Ba$_{1-x}$Sr$_x$SnO$_3$ ($x$ = 0, 0.03, 0.06, and 0.1) were measured at the Steady High Magnetic Field Facility, Hefei, China. Fig. 4 presented the Raman spectra of the parent compound BSO. With the increasing temperature, the Raman excitations merged together and the number of Raman active modes decreased (see Fig. 4(a)). Especially, a "transient state" was demonstrated as the excitations slightly converged together in the range of 69.5 - 101 K [19]. Meanwhile, the excitations at 357, 527 and 657 cm$^{-1}$ disappeared and the excitation at 409 cm$^{-1}$ appeared around 69.5 K. Since the Raman active mode should not be observed in the ideal cubic lattice structure with the space group Pm$\bar{3}$m, the Raman modes at 5.6 K suggested the lowered lattice symmetry caused by a potential lattice distortion.

The phonon dispersions and partial density of states of BSO were simulated by *phonopy* code with linear response method, Fig.4(c) and (d). The LO-TO splitting induced by electrostatic potential in nonmetals was considered. The optical modes in the frequency range of 150 - 425 cm$^{-1}$ were contributed by the distorted SnO$_6$ octahedra. The high energy scale, above 500 cm$^{-1}$, primarily related to the O-O modes. The phonon frequency 480 cm$^{-1}$ was located in the band gap and regarded as second-order Raman modes. The longitudinal (LO) and transverse (TO) optical phonon modes were selected to analyze the phonon behavior, Table 2. According to the group theory and transition rule, the phonon states transformed from Pm$\bar{3}$m to tetragonal P4/mbm[20], orthorhombic Pbnm or even rhombohedral R$\bar{3}$c could explain the observed discrepancy between the theory and experiment.[21] The lattice distortion shifted zone-center to $M_1$- or R-point from Pm$\bar{3}$m to P4/mbm or R$\bar{3}$c, respectively. To examine the Raman modes of BSO with the space group Pbnm, the harmonic lattice vibrations were assumed at 5.6 K. In the harmonic approximation with the fixed force constant, the relation of the masses and frequencies of BSO and SSO within the same structure was shown in the following equation[22]:

$$\frac{v_{Ba}}{v_{Sr}} \approx \sqrt{\frac{M_{Sr}}{M_{Ba}}} \quad (9)$$

where $\upsilon$ and M were the vibration frequency and atomic mass, respectively. Thus, it was obvious that the zone-center phonon shift to M- or R- points in FBZ and a lattice distortion from dynamic phase transition was occurred in the undoped BSO.

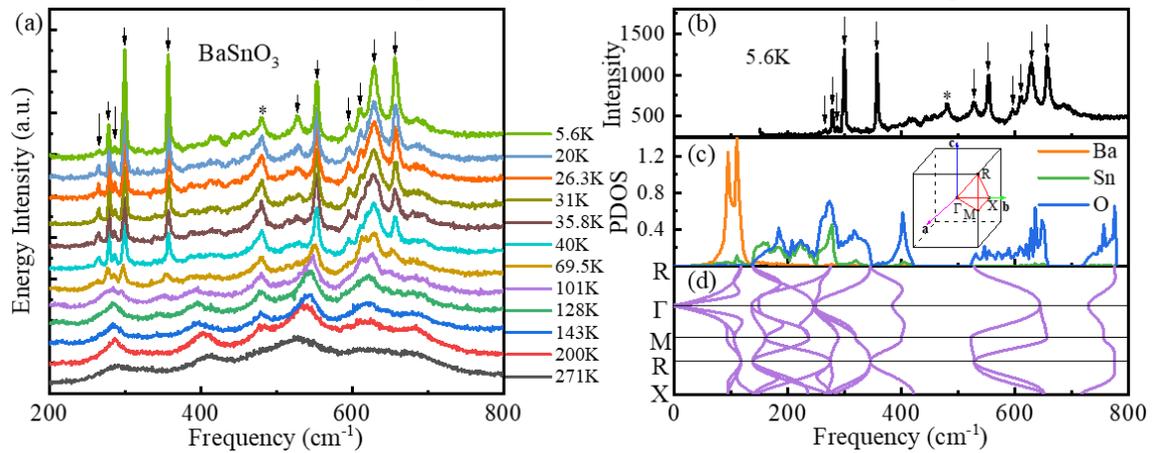

Figure 4. (a) Temperature dependence of the Raman spectrum of BaSnO$_3$. (b) The Raman spectrum of BaSnO$_3$ at 5.6 K. (c) and (d) The calculated partial density states and phonon dispersion of BaSnO$_3$, respectively. The inset in (c) indicated the high symmetry points in the FBZ.

Sr-doping affected the Sn-O and O-O vibrations indirectly by varying the environment of the SnO$_6$ octahedra, and the related effect on the lattice vibrations was obtained from the temperature-dependence of BSSOs, Fig.5. i) The Raman mode at 357 cm$^{-1}$ was the excitation at the zone boundary, *R*-point, and was responsible for the O-O bond. This mode had no significant change with low-concentration Sr doping at 5.6 K, whereas, the Sr$^{2+}$-ion substituting constructed the asymmetric coordination environment and distorted the lattice, corresponding to the phonons moving to the zone center from the zone boundary, i.e. M or R points. However, *R*-point phonons was not observed at 300 K, Fig. 5. ii) Both Raman active modes at 528 and 610 cm$^{-1}$ were the excitations of the O-O bonds. The redshift was observed at 300 K while the blueshift was presented at 5.6 K. Hence, the O-O vibration was determined by the competition of the low-fraction doping on the lattice distortion and the thermal effect on the lattice symmetry. Moreover, those broad and asymmetry excitations ruled out that Sr doping strengthened the O-O polarizability. iii) The Raman active mode at 629 cm$^{-1}$ was contributed by both Sn-O and O-O modes, and demonstrated that Sr-substitution enhanced the distortion of SnO$_6$ octahedra. Furthermore, this phonon excitation was not observed at 300 K and suggested that the lattice distortion was less at high temperature.

To further understand the lattice distortion and quantitatively describe the phase transition temperature in BSSO, the temperature-dependence of Raman shifts were presented in Fig. 5(j)-(l). Firstly, the temperature dependence of Raman shifts could be simply described[23]

$$\omega = \omega_0 - \alpha_1 T - \alpha_2 T^2 \quad (10)$$

where $\omega_0$ was the excitation at 0 K, $\alpha_1$ and $\alpha_2$ were the first- and second-order temperature coefficients, respectively.

The wavenumbers could be obtained from the combination of lattice and phonon-phonon anharmonic terms as the decay of two phonons $\omega_1$ and $\omega_2$, and three identical phonons of $\omega/3$ [24]:

$$\omega = \omega_0 + A\left(1 + \frac{1}{e^{x_1}-1} + \frac{1}{e^{x_2}-1}\right) + B\left(1 + \frac{3}{e^{x_3}-1} + \frac{3}{(e^{x_3}-1)^2}\right) \quad (11)$$

where $x_1 = \hbar\omega_1/k_B T$, $x_2 = \hbar\omega_2/k_B T$, $x_3 = \hbar\omega_3/k_B T$. The parameters A and B were the third and fourth order decay, respectively. $\hbar$, $k_B$ and $T$ denoted the Planck's constant, the Boltzmann's constant and the temperature, respectively. The frequency $\omega_0$ was obtained by Eq. (14).

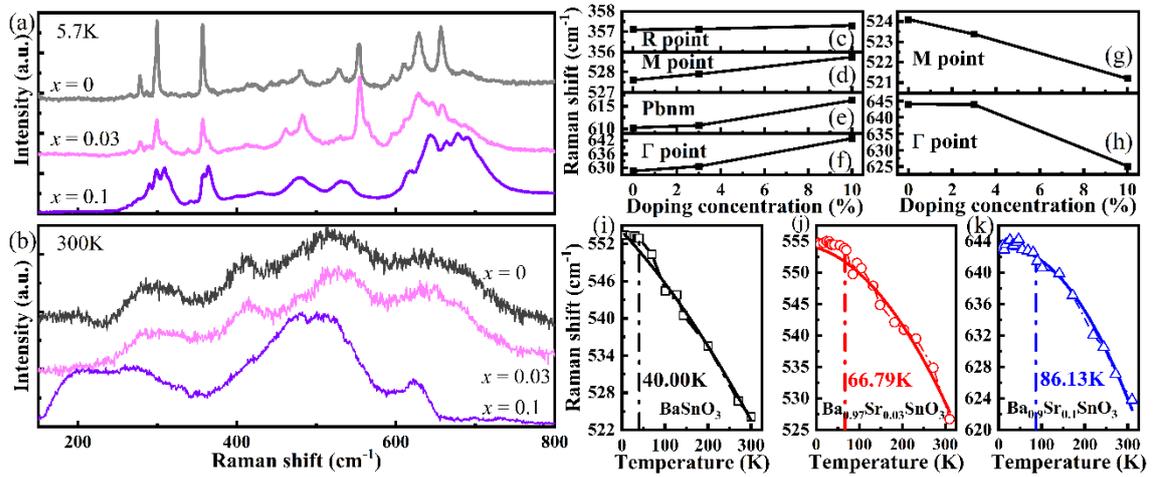

Figure 5. The Raman spectra of Ba$_{1-x}$Sr$_x$SnO$_3$ ($x$ = 0, 0.03, 0.1) at 5.7 K(a) and 300 K(b). (c)-(h) Composition-dependence of some intense frequencies for Ba$_{1-x}$Sr$_x$SnO$_3$ ($x$ = 0, 0.03, 0.1) at 5.6 K and 300 K. The temperature-dependence of the Raman excitations of (i)BaSnO$_3$, (j)Ba$_{0.97}$Sr$_{0.03}$SnO$_3$, (k)Ba$_{0.9}$Sr$_{0.1}$SnO$_3$.

The temperature-dependence of 554 cm$^{-1}$ (TO$_3$) for BSO corresponded to O-O polarization, Fig. 5(i), and an anomaly was observed around 40 K, which not only suggested a lattice distortion on the Raman active mode, but also agreed with the heat capacity of BSO, inset Fig. 2(a). In the cases of $Ba_{0.97}Sr_{0.03}SnO_3$, Fig. 5(j), the phase transition temperature increased to 66.79 K. The Raman mode at 644 cm$^{-1}$ was originated from the tortuosity of Sn-O octahedra and O-O polarization for $Ba_{0.9}Sr_{0.1}SnO_3$, Fig. 5(k), and the transition temperature was around 86.13 K. Therefore, the Sr doping strengthened the bond energy and lattice distortion to zone boundary.

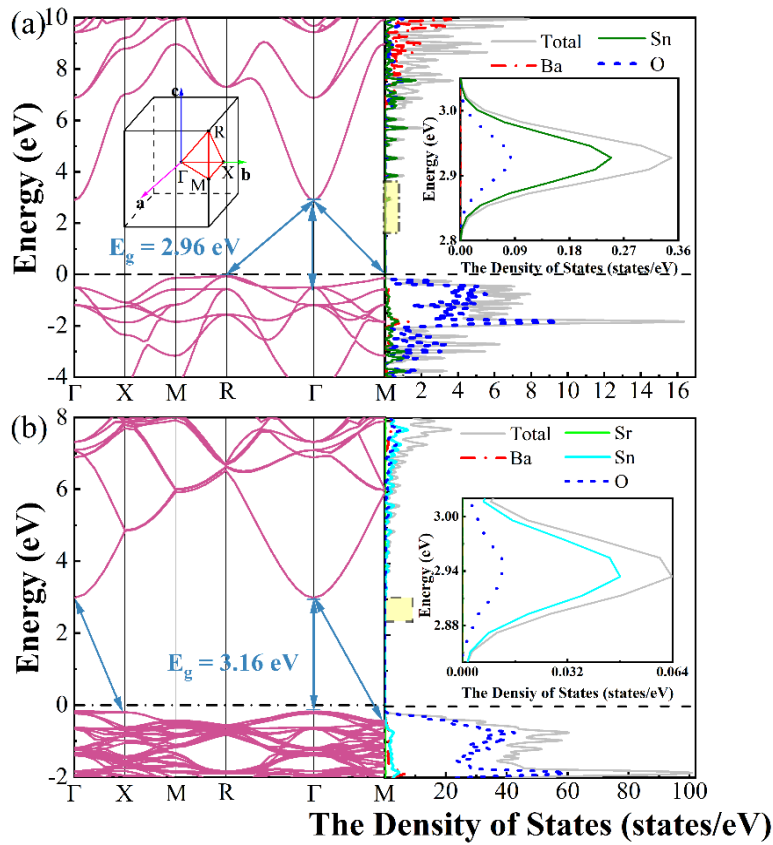

Figure 6. The band structure and total density of states of $BaSnO_3$ in unit cell (a) and $Ba_{0.875}Sr_{0.125}SnO_3$ in 2×2×2 supercell (b), respectively. Fermi levels were described by black dash line.

**Electronic structures of $Ba_{1-x}Sr_xSnO_3$(x = 0, 0.125).** The electronic properties of $Ba_{1-x}Sr_xSnO_3$(x = 0, 0.125) were discussed theoretically through the band structures, total and partial density of states using HSE06 hybrid functional calculation. The lattice parameter of the optimized structure of $BaSnO_3$ was 4.10 Å as the NPD experiment. In Fig. 6 (a), the conduction band minimum (CBM) and valence band maximum (VBM) located at Γ and R points, respectively, and the indirect band gap was 2.96 eV, which was consistent with the experiment results 2.93-3.4 eV, Table 3.[15, 25] The total density of states (TDOSs) indicated that Sn and O atoms mainly contributed to the conduction and valence band, respectively. The DOS of Ba atoms was much smaller in both bands, while the electrons located at *s* orbit primarily formed the Sn states, and the electrons dominating *p* orbits offered the foremost contribution to the O states just below Fermi energy (see in Fig. 7).

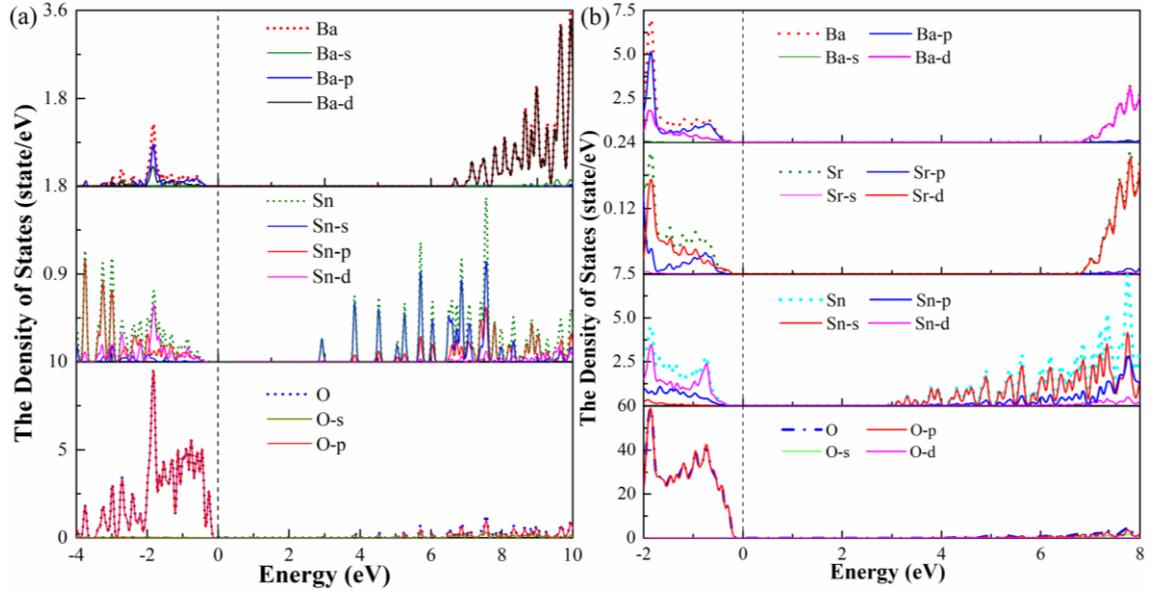

Figure 7. The partial density of states of BaSnO$_3$ in unit cell (a) and Ba$_{0.875}$Sr$_{0.125}$SnO$_3$ in 2×2×2 supercell (b), respectively. Fermi levels were described by black dash line.

For Ba$_{0.875}$Sr$_{0.125}$SnO$_3$, 2×2×2 supercell was applied and the lattice parameter was 4.08 Å. Significantly changes in the band structures of CBM and VBM were obtained at the Γ point by Sr substitution. The electron was removed from the Fermi energy by a large bandgap energy of 3.16 eV, and the indirect band gap was tuned to the direct band gap as the distorted BaSnO$_3$.[20] Hence, the Sr-doping induced the zone center to M or R points and valence electrons near the Fermi surface transferred to the Γ or X points, Table 3.

Compared to the Raman excitations with the Sr-doping, the electron-phonon coupling was clearly broken and the energy gap became large for a stable structure. This similar feature was obtained in Ba$_{0.75}$Sr$_{0.25}$SnO$_3$ and Ba$_{0.25}$Sr$_{0.75}$SnO$_3$, Table 3.[23] For SSO, the lattice is pseudo-cubic with the space group Pbnm and the electron band was back to an indirect one as 3.95 eV. Hence, we offered an effective approach to improve the optical properties of BSO and essentially illustrate the reason for the migration of electron and phonon behavior.

**Table 3. The band gaps (in eV) along different electronic excitation paths. The symbol ♦ labeled the energy gap from VBM to CBM.**

| Configuration | Γ→Γ | X→Γ | M→Γ | R→Γ |
|---|---|---|---|---|
| BaSnO$_3$ | 3.40 | 3.95 | 3.10 | 2.96♦ |
| Ba$_{0.875}$Sr$_{0.125}$SnO$_3$ | 3.16♦ | 3.19 | 3.42 | 3.53 |
| Ba$_{0.75}$Sr$_{0.25}$SnO$_3$ | 3.22♦ | 3.25 | 3.53 | 3.62 |
| Ba$_{0.25}$Sr$_{0.75}$SnO$_3$ | 3.83 | 4.00 | 3.74♦ | 3.99 |

In Fig. 6, CBM and VBM were dominated by Sn and O atoms, respectively, and O states moved away from the Fermi level to enlarge the energy gap by doping Sr-ions. To further understand the role of each orbital, PDOS of Ba$_{0.875}$Sr$_{0.125}$SnO$_3$ was simulated, Fig. 7, and the Ba and Sr states near the Fermi energy were from the electrons occupying *p* and *d* orbits, respectively. The bottom of the conduction region and top of the valence region was contributed by Sn-s and O-p orbits, respectively.

CONCLUSION

In summary, we systematically investigated the structural, thermal properties and electronic structures of Sr-doped BSO ceramics. The temperature-dependence Raman spectra revealed that the intrinsic lattice distortion in the undoped BSO activated the Raman modes and induced zone-center phonon shifted to zone boundary, finally, a potential dynamic phase transition occurred as the temperature increased. The electrons below the Fermi level located at M or R points and could interact with the moved phonons, which enhanced the anharmonic effect. Sr doping strengthened the lattice distortion and lowered the phonon symmetry. In contrast, the electron below the Fermi energy moved to Γ point and the conduction bond minimum always located at Γ point, resulting in the direct band gap. Thus, Sr doping could tune the indirect band gap to direct band gap and break the electron-phonon coupling. Finally, the gap energy became larger in order to lower energy states.

EXPERIMENTAL AND THEORETICAL DETAILS

**Synthesis of $Ba_{1-x}Sr_xSnO_3$.** The precursors of $BaCO_3$ (99.97%), $SrCO_3$ (99.9%), and $SnO_2$ (99.9%) were used to synthesize the polycrystalline samples of $Ba_{1-x}Sr_xSnO_3$ with $x = 0, 0.03, 0.06, 0.1$ via solid state reaction.

**Neutron powder diffraction.** The neutron powder diffraction (NPD) of $Ba_{1-x}Sr_xSnO_3$ were taken on BL08 Super High Resolution Powder Diffractometer time-of-flight diffractometer (SuperHRPD) at the Material and Life Science Facility of Japan Proton Accelerator Research Complex (J-PARC). The diffraction datum was collected at the various temperature ranging from 10 K to 300 K and the Rietveld refinement was performed by the Z-Rietveld software.

**Physical properties.** The heat capacity and the thermal conductivity were carried out by the Physical Property Measurement System (PPMS).

The temperature dependent Raman spectra of $Ba_{1-x}Sr_xSnO_3$ were obtained by the DXR Raman Microscope (Horiba JY T64000) with the laser of 532.1 nm and a Janis ST-500 microscopy cryostat down to 5.6 K. The scanning range was from 150-800 cm$^{-1}$ with the resolution 0.5 cm$^{-1}$ and the data were recorded by the charged-coupled device (CCD) of single exposure.

**First-principles calculation.** The calculations were performed by the first-principles analysis based on density functional theory (DFT) as implemented in the Vienna *Ab initio* Simulation Package (VASP) [26, 27]. The ion-electron interactions were approximated using the projected-augmented plane method (PAW)[28], and the local density approximation (LDA) in the form of Ceperley-Alder-Perdew-Zunger (CAPZ) parameterization[28, 29] was employed to simulate the electron-exchange correlation potentials. The plane wave basis with a cutoff energy of 500 eV was used.

To investigate the influence of Sr doping on the electronic structures of $BaSnO_3$, we selected a 2×2×2 supercell containing a whole number of 40 atoms whereas one Ba atom was substituted by one Sr atom to insightfully observe the effect of 12.5% Sr-doped concentration. The integrations over the bulk Brillouin zone of the $BaSnO_3$ unit cell and the 2×2×2 supercell of $Ba_{0.875}Sr_{0.125}SnO_3$ were calculated by the Monkhost-Pack special k meshes 9×9×9 and 7×7×7, respectively. In order to the full structural optimization, the energy and force convergence standard for each atom were set to $10^{-7}$ eV and 0.01eV/Å, respectively. We utilized the Hartree-Fock/DFT hybrid functional[30] with the range-separation parameter of 0.2 Å$^{-1}$ to obtain the accurate band structures[31-33] and the default fraction of exact exchange was 25%. The valence electronic configuration of Ba and Sr atoms was $5s^25p^66s^2$ and $4s^24p^65s^2$, respectively. $Ba^{2+}$ and $Sr^{2+}$ cations were generated in $Ba_{1-x}Sr_xSnO_3$. For Sn atom, the electrons occupying the $5s^2$ and $5p^2$ orbits were taken in valence. The valence electrons of O atom were 6 ($2s^22p^4$).

The phonon behavior of $BaSnO_3$ was calculated through *phonopy* code based on DFT. We enlarged the primitive cell to the 4×4×4 supercell for the computation of phonon dispersion, PDOSs and specific heat. The Gamma-centered k-point grids were selected as 5×5×5 and the energy convergence was set to $10^{-9}$ eV.

AUTHOR INFORMATION

Corresponding Authors

**Haidong Zhou** – *Department of Physics and Astronomy, University of Tennessee, Knoxville, Tennessee 37996-1200, USA*; Email: hzhou10@utk.edu

**Jie Ma** - *Key Laboratory of Artificial Structures and Quantum Control, School of Physics and Astronomy, Shanghai Jiao Tong University, Shanghai 200240, China*; *Wuhan National High Magnetic Field Center, Huazhong University of Science and Technology, Wuhan 430074, China*; Email: jma3@sjtu.edu.cn

Authors

**Binru Zhao** - *Key Laboratory of Artificial Structures and Quantum Control, School of Physics and Astronomy, Shanghai Jiao Tong University, Shanghai 200240, China*; Email: zebra2020@sjtu.edu.cn

**Qing Huang** - *Department of Physics and Astronomy, University of Tennessee, Knoxville, Tennessee 37996-1200, USA*; Email: qhuang11@vols.utk.edu

**Jiangtao Wu** – *Key Laboratory of Artificial Structures and Quantum Control, School of Physics and Astronomy, Shanghai Jiao Tong University, Shanghai 200240, China*; Email: biteskye@sjtu.edu.cn

**Jinlong Jiao** - *Key Laboratory of Artificial Structures and Quantum Control, School of Physics and Astronomy, Shanghai Jiao Tong University, Shanghai 200240, China*; Email: jjl.sjtu@sjtu.edu.cn

**Mingfang Shu** - *Key Laboratory of Artificial Structures and Quantum Control, School of Physics and Astronomy, Shanghai Jiao Tong University, Shanghai 200240,*

*China*; Email: shumf123@sjtu.edu.cn

**Gaoting Lin** - *Key Laboratory of Artificial Structures and Quantum Control, School of Physics and Astronomy, Shanghai Jiao Tong University, Shanghai 200240, China*; Email: doublewood@sjtu.edu.cn

**Qiyang Sun** – *Materials Science and Engineering/Department of Mechanical Engineering, University of California, Riverside, California 92521, USA*; Email: qsun026@ucr.edu

**Ranran Zhang** - *High Magnetic Field Laboratory, Chinese Academy of Sciences, Hefei 230031, China*; Email: zhangrr@hmfl.ac.cn

**Masato Hagihala** - *Institute of Materials Structure Science, High Energy Accelerator Research Organization, Tokai, Ibaraki 319-1106, Japan*; Email: hagihala@post.kek.jp

**Shuki Torri** - *Institute of Materials Structure Science, High Energy Accelerator Research Organization, Tokai, Ibaraki 319-1106, Japan*; Email: torii@post.kek.jp

**Guohua Wang** - *Key Laboratory of Artificial Structures and Quantum Control, School of Physics and Astronomy, Shanghai Jiao Tong University, Shanghai 200240, China*; Email: wangguohua190196@sjtu.edu.cn

**Qingyong Ren** - *China Spallation Neutron Source, Institute of High Energy Physics Chinese Academy of Sciences, Dongguan 523803, China*; Email: renqy@ihep.ac.cn

**Chen Li** - *Materials Science and Engineering/Department of Mechanical Engineering, University of California, Riverside, California 92521, USA*; Email: chenli@ucr.edu

**Zhe Qu** - *Anhui Province Key Laboratory of Condensed Matter Physics at Extreme Conditions, High Magnetic Field Laboratory, Hefei Institutes of Physical Sciences, Chinese Academy of Science, Hefei,Anhui 230031, China*; Email: zhequ@hmfl.ac.cn

Notes
The authors declare no competing financial interest.

ACKNOWLEDGMENTS

B.R.Z., J.T.W., J.L.J., M.F.S., G.T.L., G.H.W., Q.Y.R and J.M. acknowledge the financial support from the National Science Foundation of China (Nos. 11774223, and U2032213), the interdisciplinary program Wuhan National High Magnetic Field Center (Grant No. WHMFC 202122), Huazhong University of Science and Technology. The work at the University of Tennessee (Q.H., and H.D.Z.) was supported by the NSF with Grant No. NSF-DMR-2003117. G.T.L. thanks the projected funded by China Postdoctoral Science Foundation (Grant no. 2019M661474) and the National Science Foundation of China (12004243). The computations in this paper were run on the π 2.0 cluster supported by the Center for High Performance Computing at Shanghai Jiao Tong University.

# REFERENCES

(1) Bai, Y.; Siponkoski, T.; Peräntie, J., *et al.* Ferroelectric, pyroelectric, and piezoelectric properties of a photovoltaic perovskite oxide. *Applied Physics Letters* **2017,** 110.

(2) Bouhemadou, A.; Haddadi, K. Structural, elastic, electronic and thermal properties of the cubic perovskite-type BaSnO3. *Solid State Sciences* **2010,** 12, 630-636.

(3) Giustino, F. Electron-phonon interactions from first principles. *Reviews of Modern Physics* **2017,** 89, 015003.

(4) Peter, Y.; Cardona, M., *Fundamentals of semiconductors: physics and materials properties*. Springer Science & Business Media: 2010.

(5) Zhang, Y.; Wang, J.; Sahoo, M., *et al.* Strain-induced ferroelectricity and lattice coupling in BaSnO 3 and SrSnO 3. *Physical Chemistry Chemical Physics* **2017,** 19, 26047-26055.

(6) Shockley, W.; Bardeen, J. Energy bands and mobilities in monatomic semiconductors. *Physical Review* **1950,** 77, 407.

(7) Bardeen, J.; Shockley, W. Deformation potentials and mobilities in non-polar crystals. *Physical review* **1950,** 80, 72.

(8) Franchini, C.; Reticcioli, M.; Setvin, M., *et al.* Polarons in materials. *Nature Reviews Materials* **2021**, 1-27.

(9) Cohen, R. E. Origin of ferroelectricity in perovskite oxides. *Nature* **1992,** 358, 136-138.

(10) Ong, K. P.; Fan, X.; Subedi, A., *et al.* Transparent conducting properties of SrSnO3 and ZnSnO3. *APL Materials* **2015,** 3.

(11) Singh, D. J.; Xu, Q.; Ong, K. P. Strain effects on the band gap and optical properties of perovskite SrSnO3 and BaSnO3. *Applied Physics Letters* **2014,** 104, 011910.

(12) Wang, T.; Prakash, A.; Dong, Y., *et al.* Engineering SrSnO3 phases and electron mobility at room temperature using epitaxial strain. *ACS applied materials & interfaces* **2018,** 10, 43802-43808.

(13) Bellal, B.; Hadjarab, B.; Bouguelia, A., *et al.* Visible light photocatalytic reduction of water using SrSnO 3 sensitized by CuFeO 2. *Theoretical and Experimental Chemistry* **2009,** 45, 172-179.

(14) Shannon, R. D. Revised effective ionic radii and systematic studies of interatomic distances in halides and chalcogenides. *Acta crystallographica section A: crystal physics, diffraction, theoretical and general crystallography* **1976,** 32, 751-767.

(15) Smith, M. G.; Goodenough, J. B.; Manthiram, A., *et al.* Tin and antimony valence states in BaSn0.85Sb0.15O3−δ. *Journal of Solid State Chemistry* **1992,** 98, 181-186.

(16) Petersen, A.; Bhattacharya, S.; Tritt, T. M., *et al.* Critical analysis of lattice thermal conductivity of half-Heusler alloys using variations of Callaway model. *Journal of Applied Physics* **2015,** 117, 035706.

(17) Roufosse, M. C.; Klemens, P. G. Lattice thermal conductivity of minerals at high temperatures. *Journal of Geophysical Research* **1974,** 79, 703-705.

(18) Yadav, K.; Singh, S.; Muthuswamy, O., *et al.* Unravelling the phonon scattering mechanism in Half-Heusler alloys ZrCo1-xIrxSb (x= 0, 0.1, and 0.25). *arXiv preprint arXiv:2107.11567* **2021**.

(19) Xu, C.; Wang, C.; Yu, J., *et al.* Structure and optical properties of Er-doped CaO-Al2O3 (Ga2O3) glasses fabricated by aerodynamic levitation. *Journal of the American Ceramic Society* **2017,** 100, 2852-2858.

(20)     Liu, H.-R.; Yang, J.-H.; Xiang, H. J., *et al.* Origin of the superior conductivity of perovskite Ba(Sr)SnO3. *Applied Physics Letters* **2013,** 102.

(21)     Smirnova, I. S.; Bazhenov, A. V.; Fursova, T. N., *et al.* IR-active optical phonons in Pnma-1, Pnma-2 and phases of. *Physica B: Condensed Matter* **2008,** 403, 3896-3902.

(22)     Kreisel, J.; Glazer, A.; Jones, G., *et al.* An x-ray diffraction and Raman spectroscopy investigation of A-site substituted perovskite compounds: the (Na1-xKx) 0.5 Bi0. 5TiO3 (0 x1) solid solution. *Journal of Physics: Condensed Matter* **2000,** 12, 3267.

(23)     Li, W.; Shen, Z.; Feng, Z., *et al.* Temperature dependence of Raman scattering in hexagonal gallium nitride films. *Journal of Applied Physics* **2000,** 87, 3332-3337.

(24)     Balkanski, M.; Wallis, R. F.; Haro, E. Anharmonic effects in light scattering due to optical phonons in silicon. *Physical Review B* **1983,** 28, 1928-1934.

(25)     Kim, H. J.; Kim, J.; Kim, T. H., *et al.* Indications of strong neutral impurity scattering in Ba(Sn,Sb)O3single crystals. *Physical Review B* **2013,** 88.

(26)     Zhang, Y.; Wu, X. Vanadium sulfide nanoribbons: Electronic and magnetic properties. *Physics Letters A* **2013,** 377, 3154-3157.

(27)     Kresse, G.; Furthmüller, J. Efficient iterative schemes for ab initio total-energy calculations using a plane-wave basis set. *Physical review B* **1996,** 54, 11169.

(28)     Perdew, J. P.; Wang, Y. Accurate and simple analytic representation of the electron-gas correlation energy. *Physical review B* **1992,** 45, 13244.

(29)     Ceperley, D. M.; Alder, B. J. Ground state of the electron gas by a stochastic method. *Physical review letters* **1980,** 45, 566.

(30)     Heyd, J.; Scuseria, G. E.; Ernzerhof, M. Hybrid functionals based on a screened Coulomb potential. *The Journal of chemical physics* **2003,** 118, 8207-8215.

(31)     Paier, J.; Marsman, M.; Hummer, K., *et al.* Screened hybrid density functionals applied to solids. *The Journal of chemical physics* **2006,** 124, 154709.

(32)     Bjaalie, L.; Himmetoglu, B.; Weston, L., *et al.* Oxide interfaces for novel electronic applications. *New Journal of Physics* **2014,** 16, 025005.

(33)     Grüneis, A.; Kresse, G.; Hinuma, Y., *et al.* Ionization potentials of solids: the importance of vertex corrections. *Physical review letters* **2014,** 112, 096401.